%% file: main.tex
\documentclass[paper]{ieice}
\usepackage{graphicx} % Required for inserting images
\usepackage{xspace}
\usepackage{xcolor}
\usepackage{url}
\newcommand{\etal}{\xspace{ et~al.}\xspace}
\newcommand{\projectCount}{36,464}

\title{Myth: The loss of core developers is a critical issue for OSS communities}
\authorlist{%
\authorentry{Olivier Nourry}{}{}
\authorentry{Masanari Kondo}{}{}
\authorentry{Shinobu Saito}{}{}
\authorentry{Yukako Iimura}{}{}
\authorentry{Naoyasu Ubayashi}{}{}
\authorentry{Yasutaka Kamei}{}{}
}
\date{October 2024}

\begin{document}

\maketitle

\textbf{[Background]} Throughout their lifetime, open-source software systems will naturally attract new contributors and lose existing contributors. Not all OSS contributors are equal, however, as some contributors within a project possess significant knowledge and expertise of the codebase (i.e., core developers). When investigating a project’s ability to attract new contributors and how often a project loses contributors, it is therefore important to take into account the expertise of the contributors. \textbf{[Goal]} Since core developers are vital to a project’s longevity, we therefore aim to find out: can OSS projects attract new core developers and how often do OSS projects lose core developers?
\textbf{[Results]} To investigate core developer contribution patterns, we calculate the truck factor (or bus factor) of over 36,000 OSS projects to investigate how often TF developers join or abandon OSS projects. We find that 89\% of our studied projects have experienced losing their core development team at least once. Our results also show that in 70\% of cases, this project abandonment happens within the first three years of a project’s life. We also find that most OSS projects rely on a single core developer to maintain development activities. Finally, we find that only 27\% of projects that were abandoned were able to attract at least one new TF developer.
\textbf{[Discussion]}
Our analysis shows that it is not uncommon for OSS projects to lose their initial core development team. This is likely due to most OSS project relying on a single core developer to maintain development activities. The first year of development is critical for OSS projects since this is where they are most at risk of losing their core developer(s). Additionally, projects that lose their core developer(s) early seem less likely to survive this event than projects that lost their core developers later on during their life. This could be in part due to older projects having more time to build a community and be useful to more users who can revive a project for their own personal use should the original core developer(s) abandon the project. 

\section{Introduction}
\input{introduction}

\label{sec:introduction}

\section{Background and Related Work}
\input{background}
\label{sec:background}

\section{Methodology}
\input{methodology}
\label{sec:methodology}

\section{Results}
\input{results}
\label{sec:results}

\section{Conclusion}
\input{conclusion}
\label{sec:conclusion}

\newpage
%\bibliographystyle{plain}
%\bibliography{myrefs}
\bibliographystyle{ieicetr}
\bibliography{bibliography}
\end{document}

%% file: introduction.tex
The Open-source software ecosystem has become one of the most important pillar of software development over time. Today, almost every software company in the world uses open-source software to some extent. Open-source projects such as Linux, kubernetes, Docker, Tensorflow, Apache HTTP Server have revolutionized the way software development is conducted and are used by the entire software development world. To maintain software development activities, these critical open-source projects all depend on open-source contributors to keep the project active and add new features. Specifically, these projects often tend to rely on a few core developers which have been actively working on these projects for years and are very knowledgeable about the codebase. Consequently, the loss of expertise incurred by the loss of core developers (or turnover) can have significant impact on a project's development and the overall productivity of the development team. To get a better understanding of open-source project abandonment, researchers have therefore tried to conduct studies to study open-source core developers' development patterns \cite{Soheal19_iaffaldano,ICSE24_jamieson,IWCS14_hannebauer,CHASE13_steinmacher}.

One common metric to identify these core developers in open-source projects is called the truck factor (or bus factor). The truck factor metric refers to the amount of developers that can stop contributing (or get hit by a truck) before a project is at risk of dying. When all truck factor developers (or core developers) quit a project we refer to this event as a Truck Factor Developer Detachment (TFDD). Conversely, if a project has experienced a TFDD and is currently inactive or at risk of dying but is able to attract a new core developer, we define this event as a project survival. Using these metrics, researchers have been able to study the development activity of core developers in open-source projects \cite{WETSoM2011_torchiano,ICPC2016_Avelino,ESEM2019_Avelino}. Due to the heavy computational cost of calculating the truck factor, most studies so far have been conducted with less than 50 open-source projects. To the best of our knowledge, as of 2024, only one study (led by Avelino \etal \cite{ESEM2019_Avelino}) has used over 1,000 projects to study developers' software development activities. In this study, Avelino \etal compute the truck factor in 2,000 to study the abandonment of open-source projects by open-source contributors. While 2,000 projects is a significant leap over previous studies that used the truck factor metric, 2,000 projects is still too few projects to get an overview of the entire ecosystem and truly understand how common it is for core developers to abandon open-source projects. Additionally, because the study conducted by Avelino \etal focuses exclusively on very popular projects with high number of stars, this previous work does not reflect the reality of maintaining a core development team for the average (smaller) open-source project.

To address this limitation, we therefore decided to conduct the first large scale empirical study using the truck factor by replicating Avelino \etal's study using \projectCount projects. In this work, we therefore aim to address the following research questions.

\begin{itemize}
    \item RQ1) How common are TFDDs in GitHub projects?
    \item RQ2) How often do open-source projects survive a TFDD?
    \item RQ3) How do surviving projects differ from non-surviving ones?
\end{itemize}

%% file: background.tex
\subsection{Studies on developers' contribution patterns}
Due to how critical open-source projects are to software development, some work has already been conducted to study aspects of project sustainability and developer activity in open-source projects. 

Ferreira \etal \cite{SBES2020_Ferreira} investigated the turnover of core developers in 174 open-source projects and found that there was significant developer turnover in the studied projects. From their analysis, they found that larger projects and projects that were owned by an organization both showed high rates of developer turnovers. Their results also show that projects with higher turnover tend to be slower at fixing bugs and addressing issues. 

Lin \etal \cite{ICGSE2017_lin} also studied developer turnover in 5 large industrial projects. Their results show that developers with higher ownership of the codebase tend to be more likely to stay than developers that mostly work on files created by other developers. They also find that developers that work on the source code tend to be part of a project for longer than developers that work mostly on documentation.

Other aspects of open-source contributors' development activities have also been studied. Qiu \etal \cite{CSCW2019_qiu}interviewed 15 open-source contributors to understand how open-source developers choose a project to contribute. From these interviews, they then quantitatively measure 11 factors in 9,977 projects and show that open-source developerse are less likely to contribute to projects that have strict contribution guidelines.

\subsection{Truck factor studies}
The concept of truck factor (or bus factor) was first used in the context of software engineering at the start of the millennium and was defined as the number of developers that need to stop contributing (or get hit by a truck/bus) for a project to be at risk of dying\cite{ESEM10_zazworka,PP2002_williams}. Over time, several implementations and algorithms have been proposed to calculate the truck factor \cite{ICPC2017_Ferreira,ESEM2010_Ricca,ICPC2016_Avelino,SEIP2022_Jabrayilzade}. As of 2024, multiple studies have used this metric to investigate the activities of core developers in open-source projects.

In 2010, Ricca \etal proposed one of the earliest implementation of the truck factor in the context of software engineering. Using their tool, they calculated the truck factor of 20 open-source projects using different threshold and found that projects typically rely on few truck factor developers to keep development activities going. Torchiano \etal \cite{WETSoM2011_torchiano} also measured the truck factor in 20 open-source project in their 2011 study where they tried to calculate the theoretical maximum truck factor value. Their analysis show similar patterns as Ricca \etal's results where projects seem to rely on very few core developers to maintain development activities.

Calefato \etal used the truck factor to study the abandonment of open-source projects by developers. In their study, they proposed a method to detect which developers have abandoned open-source projects and validated their approach with real open-source developers. Using their approach, they then studied developer abandonment in 18 open-source projects. Their results show that all open-source core developers take at least one break from open-source contributions and that 45\% of them will completely disengage from contributing to an open-source project for at least one year. Their study also shows that developers have between 35\% and 55\% chance of returning to an open-source project after abandoning the project.

For our study, since we aiming to conduct a large scale empirical study, we needed an implementation that was reliable but also that could scale well with large projects that have dozens or hundreds of contributors. We therefore decided to use Avelino \etal's \cite{ICPC2016_Avelino} implementation because since it proved to be able to handle the analysis of 2,000 repositories in Avelino \etal's 2019 study \cite{ESEM2019_Avelino}. To ensure that our truck factor measurements were reliable and reproducible, we also decided to use an openly available (on GitHub\footnote{https://github.com/aserg-ufmg/truck-factor}) implementation of Avelino \etal's truck factor algorithm rather than re-implementing our own version of the algorithm.

%% file: methodology.tex
\textbf{Dataset selection and filtering.} To find open-source projects, we first used the publicly available libraries.IO~\cite{librariesIO_dataset} dataset which contained the names of over 37.7 million open-source source projects along with other metrics such as where the project repositories are hosted, when the projects were created, how many stars each project has, and several other metrics. From this large dataset, we then applied a set of filters with the goal of keeping as many projects as possible while minimizing the chances of investigating toy projects. Additionally, because the truck factor is calculated on a yearly basis, our filters needed to ensure that the remaining projects had enough development history to calculate the truck factor. Our filtering criteria were therefore as follow: each project had to have a minimum of 20 stars, 10 contributors, could not be a fork, had to be hosted on GitHub, and needed at least two years of development experience (i.e., a project created in 2024 was not elligible). After applying these filters, \projectCount projects remained and were used for our study.\newline

\noindent\textbf{Data mining.} To calculate the yearly truck factor in each project, we first extracted the creation date of all repositories in our dataset. From that initial creation date, we used the \emph{git checkout} command to jumped ahead in each project's development history one year at a time. During each jump, we executed the truck factor calculation tool\footnote{https://github.com/aserg-ufmg/truck-factor} which would calculate the commit and file information of a project, determine the main programming language of the project then calculate the number of truck factor developers. Following the original paper's methodology we also data mined the name and emails of all contributors in each project to find similar names or email addresses and map them to a single entity/developer. This process was done in order to avoid cases where a developer had multiple accounts or would contribute to a GitHub repository using a different account.

To compare repositories that survived TFDDs and those that did not (RQ3), we used the official GitHub API to mine the number of commits, the number of contributors, the number of files and the age of the studied repositories. For this part, we also used the GitHub API to find out the name of each repository's main branch since we only wanted to calculate the truck factor based on the contributions pushed to the main branch (not to development branches).

\noindent\textbf{Data analysis.} After the data mining process, we then aimed to identify instances of Truck Factor Developer Detachment (TFDD) in our studied projects. To find TFDDs, we once again started from the creation date of a repository and jumped one year a time to find the date of the last commit of each developer during that year. For a given year, if a developer had not contributed (had no commits) for at least a year, we considered that this developer had abandoned the project. To identify truck factor developer detachment, we therefore used our truck factor data to identify truck factor developers then looked at the state (active/abandon) of each of these developers and flagged a project as TFDD whenever, all truck factor developers had abandoned the project. 

Using the truck factor data, the TFDD data, and the repository data, we then proceeded with the analysis to answer our research questions. To answer RQ1, we first calculated the number of projects that experienced a TFDD and how many TFDD each project experienced. To understand when open-source projects are most at risk of dying, we then calculated during which year TFDDs happened in our projects that experienced a TFDD. Additionally, we also summed up the number of TFDDs each year to calculate the cumulative percentage of TFDD year after year. Finally, we calculated how many truck factor developers our projects have to better understand how fragile (i.e., a single core developer) or robust (i.e., many core developers) open-source projects are in real world scenarios.

Next, we calculated how many of the projects that experienced a TFDD were able to survive (i.e., attract a new core developer). Additionally, we also calculated how many developers were involved with the survival of the studied projects (i.e., how many new core developers were involved with reviving the project). 

Lastly, we calculated the number of commits, the number of files, the number of contributors and the age of each project (in days) to visualize the difference between projects that survive a TFDD and those that do not. 

%% file: results.tex
\subsection{RQ1) How common are TFDDs in GitHub
projects?}
\textbf{From our \projectCount studied projects, we find that 32,689 (89.65\%) projects have faced at least one TFDD throughout their lifetime.} Calculating the number of TFDD that each project experienced, we then find that 25,642 projects experienced only a single TFDD, 6,102 projects experienced two TFDDs, 861 projects experienced three TFDDs, and 76 projects projects experienced four or more TFDDs during development.

\begin{figure}[h]
\centering
\caption{Year during which projects faced TFDDs (from the initial creation of the project).}
\includegraphics[width=0.85\columnwidth]{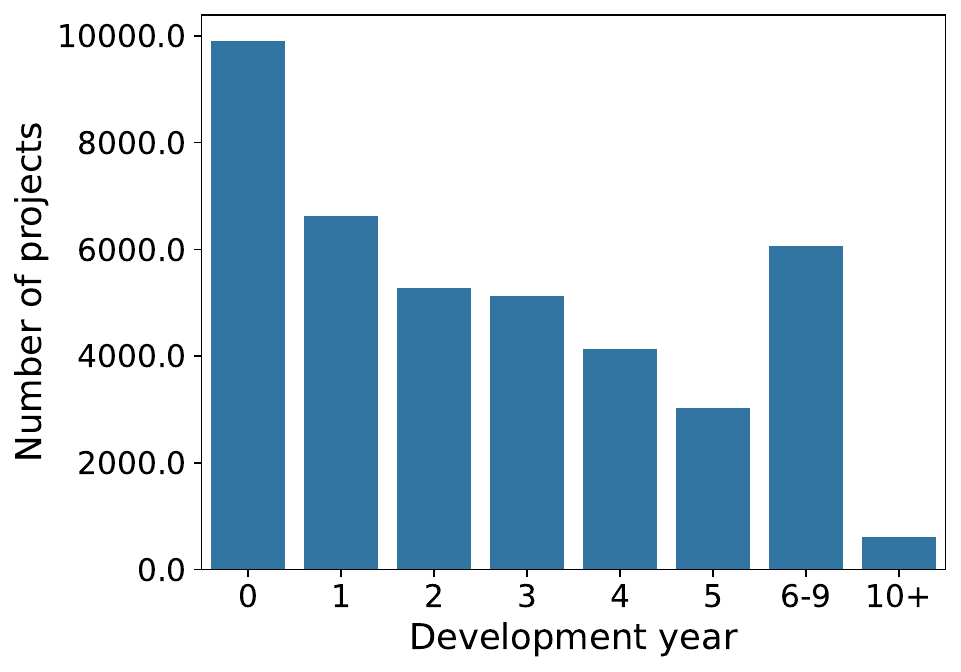}
\label{fig:tfdd_year}
\end{figure}

\textbf{Most TFDDs happen within the first year of development.} Figure \ref{fig:tfdd_year} shows during what year our studied projects experienced their first TFDD. As Figure \ref{fig:tfdd_year} shows, there are significantly more TFDDs in the first year of development with a gradual decrease during each subsequent year. 

\begin{figure}[h]
\centering
\caption{Cumulative percentage of TFDDs that happen within N years.}
\includegraphics[width=0.85\columnwidth]{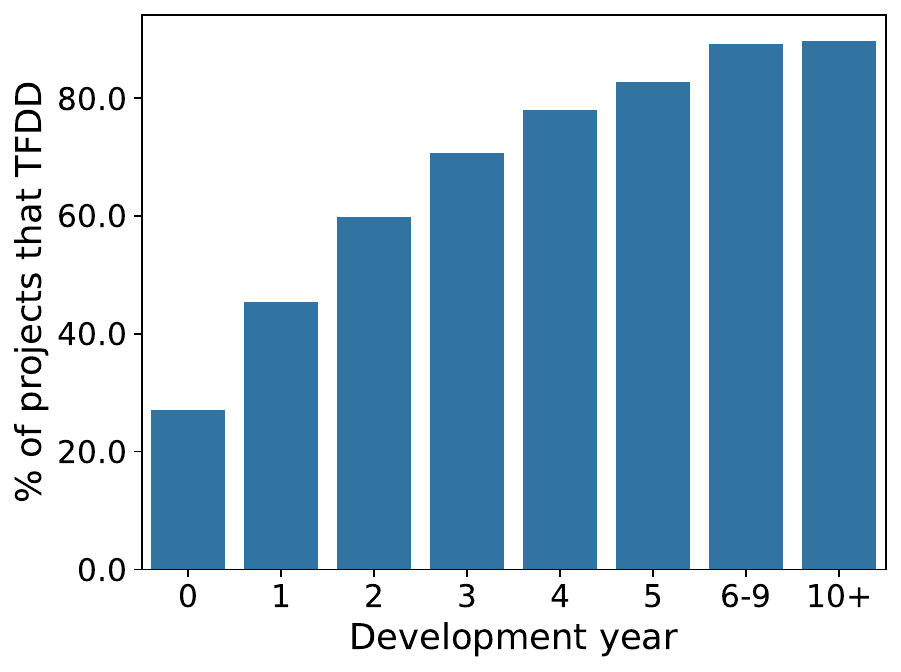}
\label{fig:cumul_project_tfdd}
\end{figure}

Figure \ref{fig:cumul_project_tfdd} shows the cumulative percentage of projects that have experienced a TFDD each year. Our results show that for projects who do experience a TFDD, 70\% of them will TFDD within the first three years, 78\% within the first four years, and 82\% with the first five years.

\begin{figure}[h]
\centering
\caption{Number of TF developers at the time of TFDD.}
\includegraphics[width=0.85\columnwidth]{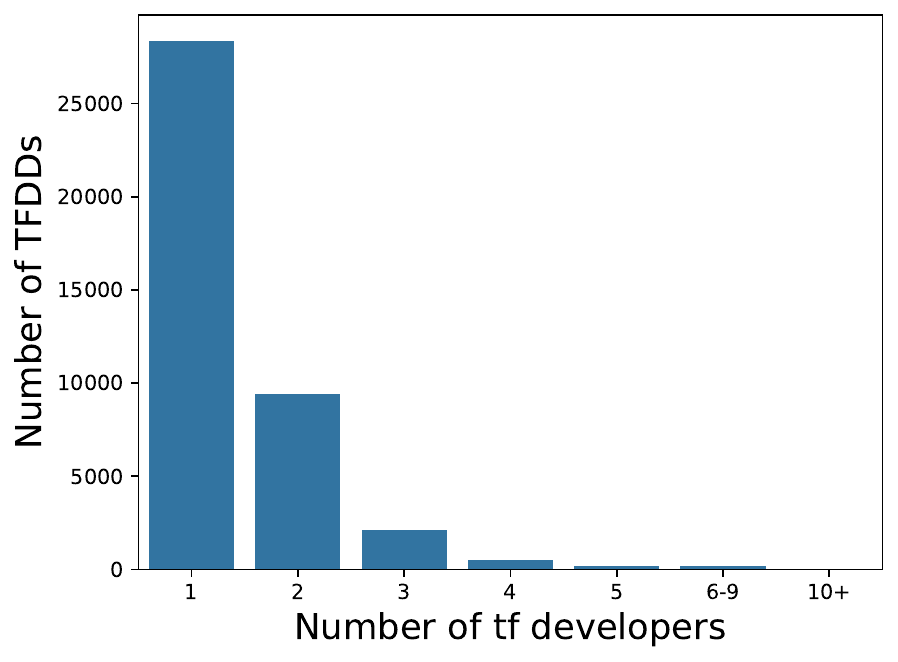}
\label{fig:tf_dev}
\end{figure}

Finally, Figure \ref{fig:tf_dev} shows the number of truck factor developers involved with our studied projects at the time of TFDD. As our results show, open-source projects relying on a single core developer to keep development activities active seem to be a common situation in the GitHub ecosystem.

\subsection{RQ2) How often do open-source projects survive a TFDD?}

\textbf{Most projects do not survive a TFDD}. Calculating the number of projects that survived a TFDD, we find that only 8,849 (27,07\%) of projects that faced a TFDD were able to survive and attract new core developers to continue development activities. From these 8,849 projects, we count 10,518 project survivals indicating that some projects were able to survive more than one TFDD. From these 10,518 TFDD survivals, we then calculated the number of developers involved with each survival. Our results shows that for 10,124 of the survivals, only one developer was involved, two developers were involved for 369 survivals and in 25 cases three developers or more took part in the project's survival.

\subsection{RQ3) How do surviving projects differ from non-surviving ones?}
\begin{figure}[h]
\centering
\caption{Number of commits, files, contributors, and project age (in days) at the time of TFDD between surviving and non-surviving projects.}
\includegraphics[width=0.85\columnwidth]{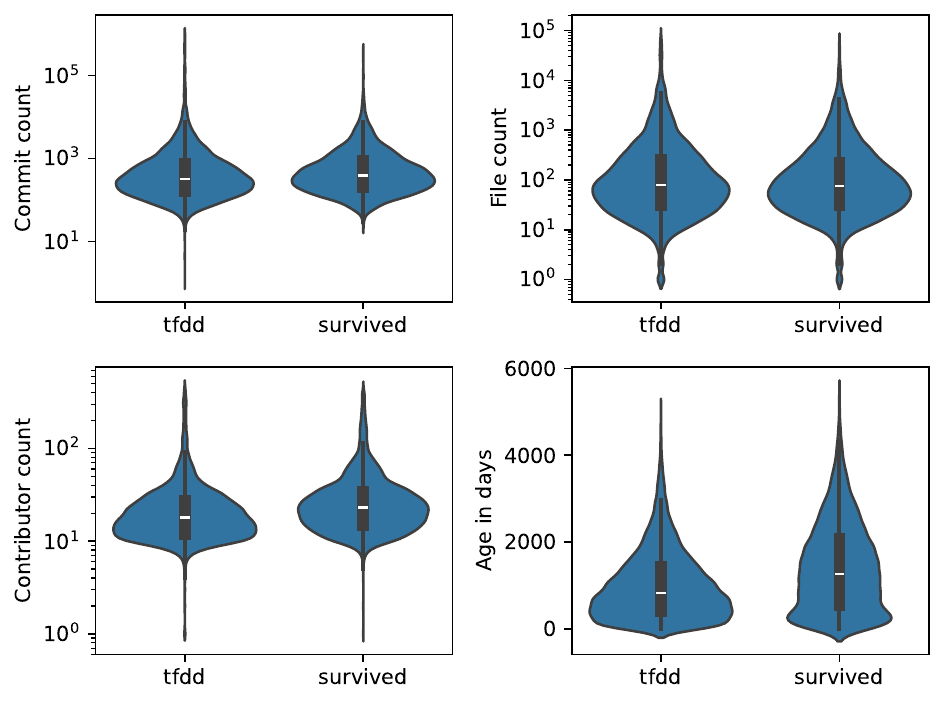}
\label{fig:rq3}
\end{figure}

\textbf{Surviving projects show more development activity than non-surviving projects across all studied metrics.} Figure \ref{fig:rq3} shows the results obtained from calculating general project metrics across all projects at the time of TFDD between projects that survived a TFDD versus projects that did not survive a TFDD. From Figure \ref{fig:rq3}, we find that projects that surviving projects have more commits and more contributors at the time of TFDD but less files than non-surviving projects. We also find that surviving projects tend to be older at the time of TFDD than non-surviving project. This indicates that more mature (older) projects are more likely to attract new developers after facing a TFDD. To ensure the statistical significance of our results, we then conduct a Mann-whitney test and find that all four studied metrics have a p-value $<$ 0.05. 

%% file: conclusion.tex
In this study, we investigated the activity of core open-source developers. Our results show that open-source projects are most at risk of getting abandoned at the start of the project's lifetime. Additionally, we also find that OSS projects often rely on a single developer to maintain development activities.